\documentclass[linenumbers]{aastex701}
\usepackage{float}

\usepackage{rotating}

\usepackage{amsmath}

\usepackage{soul}
\usepackage{silence}
\usepackage{perpage}
\MakePerPage{footnote}

\DeclareRobustCommand{\ion}[2]{%
\relax\ifmmode
\ifx\testbx\f@series
{\mathbf{#1\,\mathsc{#2}}}\else
{\mathrm{#1\,\mathsc{#2}}}\fi
\else\textup{#1\,{\mdseries\textsc{#2}}}%
\fi}

\begin{document}

\title{Unveiling Unprecedented Fine Structure in Coronal Flare Loops with the DKIST}

\author[0000-0002-3229-1848]{Cole A Tamburri}
\affiliation{DKIST Ambassador}
\affiliation{National Solar Observatory, University of Colorado Boulder, 3665 Discovery Drive, Boulder, CO 80303, USA}
\affiliation{Laboratory for Atmospheric and Space Physics, University of Colorado Boulder, 3665 Discovery Drive, Boulder, CO 80303, USA.}
\affiliation{Department of Astrophysical and Planetary Sciences, University of Colorado Boulder, 2000 Colorado Ave, CO 80305, USA}
\email{cole.tamburri@colorado.edu}

\author[0000-0001-8975-7605]{Maria D Kazachenko}
\affiliation{National Solar Observatory, University of Colorado Boulder, 3665 Discovery Drive, Boulder, CO 80303, USA}
\affiliation{Department of Astrophysical and Planetary Sciences, University of Colorado Boulder, 2000 Colorado Ave, CO 80305, USA}
\affiliation{Laboratory for Atmospheric and Space Physics, University of Colorado Boulder, 3665 Discovery Drive, Boulder, CO 80303, USA.}
\email{maria.kazachenko@lasp.colorado.edu}

\author[0000-0002-6116-7301]{Gianna Cauzzi}
\affiliation{National Solar Observatory, University of Colorado Boulder, 3665 Discovery Drive, Boulder, CO 80303, USA}
\email{gianna.cauzzi@nso.edu}

\author[0000-0003-1325-6649]{Adam F Kowalski}
\affiliation{National Solar Observatory, University of Colorado Boulder, 3665 Discovery Drive, Boulder, CO 80303, USA}
\affiliation{Department of Astrophysical and Planetary Sciences, University of Colorado Boulder, 2000 Colorado Ave, CO 80305, USA}
\affiliation{Laboratory for Atmospheric and Space Physics, University of Colorado Boulder, 3665 Discovery Drive, Boulder, CO 80303, USA.}
\email{adam.f.kowalski@colorado.edu}

\author[0000-0001-9726-0738]{Ryan French}
\affiliation{Laboratory for Atmospheric and Space Physics, University of Colorado Boulder, 3665 Discovery Drive, Boulder, CO 80303, USA.}
\email{ryan.french@lasp.colorado.edu}

\author[0000-0003-4065-0078]{Rahul Yadav}
\affiliation{National Solar Observatory, University of Colorado Boulder, 3665 Discovery Drive, Boulder, CO 80303, USA}
\affiliation{Department of Astrophysical and Planetary Sciences, University of Colorado Boulder, 2000 Colorado Ave, CO 80305, USA}
\affiliation{Laboratory for Atmospheric and Space Physics, University of Colorado Boulder, 3665 Discovery Drive, Boulder, CO 80303, USA.}
\email{rahul.yadav@lasp.colorado.edu}

\author[0000-0002-6478-3281]{Caroline L Evans}
\affiliation{National Solar Observatory, University of Colorado Boulder, 3665 Discovery Drive, Boulder, CO 80303, USA}
\affiliation{Department of Astrophysical and Planetary Sciences, University of Colorado Boulder, 2000 Colorado Ave, CO 80305, USA}
\affiliation{Cooperative Institute for Research in Environmental Sciences, 216 UCB Boulder, CO, 80309, USA}
\email{caev6801@colorado.edu}

\author[0000-0002-0412-0849]{Yuta Notsu}
\affiliation{National Solar Observatory, University of Colorado Boulder, 3665 Discovery Drive, Boulder, CO 80303, USA}
\affiliation{Laboratory for Atmospheric and Space Physics, University of Colorado Boulder, 3665 Discovery Drive, Boulder, CO 80303, USA.}
\affiliation{Department of Astrophysical and Planetary Sciences, University of Colorado Boulder, 2000 Colorado Ave, CO 80305, USA}
\email{yuta.notsu@colorado.edu}

\author[0000-0003-1597-0184]{Marcel F Corchado-Albelo}
\affiliation{DKIST Ambassador}
\affiliation{National Solar Observatory, University of Colorado Boulder, 3665 Discovery Drive, Boulder, CO 80303, USA}
\affiliation{Laboratory for Atmospheric and Space Physics, University of Colorado Boulder, 3665 Discovery Drive, Boulder, CO 80303, USA.}
\affiliation{Department of Astrophysical and Planetary Sciences, University of Colorado Boulder, 2000 Colorado Ave, CO 80305, USA}
\email{marcel.corchado@lasp.colorado.edu}

\author[0000-0002-6116-7301]{Kevin P. Reardon}
\affiliation{National Solar Observatory, University of Colorado Boulder, 3665 Discovery Drive, Boulder, CO 80303, USA}
\email{kevin.reardon@colorado.edu}
\affiliation{Department of Astrophysical and Planetary Sciences, University of Colorado Boulder, 2000 Colorado Ave, CO 80305, USA}

\author[0000-0003-3147-8026]{Alexandra Tritschler}
\affiliation{National Solar Observatory, University of Colorado Boulder, 3665 Discovery Drive, Boulder, CO 80303, USA}
\email{ali@nso.edu}

\begin{abstract}

We present the highest-resolution H$\alpha$ observations of a solar flare to date, collected during the decay phase of an X1.3-class flare on 8 August 2024 at 20:12 UT.  Observations with the Visible Broadband Imager at the National Science Foundation's (NSF) Daniel K. Inouye Solar Telescope reveal dark coronal loop strands at unprecedented spatial resolution in the flare arcade above highly structured chromospheric flare ribbons.  After surveying the $20$ best-seeing images, we calculate a mean loop width near the top of the arcade of $48.2\;km$, with a minimum loop width of $\sim$$21\;km$ and distribution mode of $\sim$$43\;km$.  The distributions of loop widths observed by the DKIST in our study are often symmetric about the mean loop width.  This is initial evidence that the DKIST may be capable of resolving the fundamental scale of coronal loops, although further investigation is required to confirm this result.  We demonstrate that the resolving power of the DKIST represents a significant step towards advancing modern flare models and our understanding of fine structure in the coronal magnetic field.

\end{abstract}


\section{Introduction} \label{sec:intro}

Given our proximity to the Sun, solar flares are the most well-observed high-energy astrophysical phenomena, with a wide range of electromagnetic spectral signatures across the different layers of the solar atmosphere.  According to the standard flare CSHKP model \citep{carmichael1963,sturrock1966,hirayama1974,kopp_pneuman1976}, reconnection in the solar corona leads to the release of magnetic free energy, precipitation of particles into the chromosphere (e.g. \citealt{shibata_magara2011,kontar2017,kowalski2024}), and often release of a coronal mass ejection (CME) into the heliosphere (e.g. \citealt{kazachenko2023}). Beneath the reconnection site is the post-eruption arcade of coronal flare loops (e.g. \citealt{Svestka1976_rcc,svestka1987,gallagher_2003,Kazachenko2022r}).  Following the convention of \cite{qiu2023}, we refer to the loops that comprise the arcade as ``post-reconnection flare loops" (PRFLs).\footnote{In the past, these have been referred to as ``post-flare loops" (e.g. \citealt{warren2018}).  We use ``post-reconnection flare loops" in this work to acknowledge that the flare continues to progress during the observations studied, but that the loops are formed as a result of magnetic reconnection associated with the flare.}  These PRFLs are filled with hot plasma ($T> 10^7 K$) evaporated from the chromospheric deposition sites, and emit at progressively cooler temperatures as the gradual flare decay phase progresses. Eventually, these loops often emit in H$\alpha$ during the flare decay phase.  From a modeling perspective the PRFL properties are defined by details of the chromospheric energy deposition, the structure of the three-dimensional magnetic field \citep{aulanier2012,aulanier2013,janvier2013,dahlin2016,dahlin2021,qiu2023,tamburri2024}, and the cross-sectional geometry of coronal loops (e.g. \citealt{klimchuk2001,mikic_2013,klimchuk2020,reep2022}).

Modern hydrodynamic (HD) and radiative-hydrodynamic (RHD) simulations demonstrate the sensitivity of flare and coronal loop evolution to the details of 3D loop geometry. The implications of loop geometry are even important to the evolution of beam-heated flare atmospheres synthesized with the 1D RADYN code \citep{carlsson1992,carlsson1995,allred2005,allred2015}.  For example, a positive gradient in flare loop cross-sectional width along the loop, required to conserve magnetic flux higher in the corona, changes the pitch angle of electrons accelerated into the chromosphere by creating a magnetic mirror within which field-aligned electrons are trapped \citep{kowalski2024}.  This phenomenon is also relevant to the recently-reported delay in signatures of particle precipitation in solar flares (e.g. \citealt{krucker2020b,chen2024}).  Similarly, the HYDrodynamics and RADiation Code (HYDRAD, \citealt{bradshaw2003,reep2013,bradshaw2016,reep2018b}) has been used in recent years to investigate the cooling times of PRFLs (e.g. \citealt{reep2020,reep2022,reep2024}).  These works demonstrate that loop expansion higher in the corona, variable cross-sectional geometry, and other parameters have a large impact on the draining of loops and spectral signatures associated with the coronal flare response.


High-resolution solar observations are crucial to determining input variables for the models described above.  Specifically, to constrain loop geometry, it is useful to resolve the smallest coronal ``strands" that comprise the arcade.  Elementary loops are theorized to have a wide range of cross-sectional widths between $10-200\;km$ (e.g. \citealt{Beveridge2003,peter2013}).  These values are smaller than the resolving power of most modern telescopes. Using flare observations from the $1.6$-meter Goode Solar Telescope at the Big Bear Solar Observatory (BBSO/GST, which has a diffraction limit of $\sim$$61\;km$ in H$\alpha$), \cite{jing2016} reported very small flare loop widths (ranging from $110\;km$ to $161\;km$).  \cite{schmidt2025} performed a similar study with GST data of an off-limb flare arcade and found a median loop width of $\sim$$107\;km$.  \cite{kuridze2013} observe loops at a resolution of $\sim$$150\;km$ with the Rapid Oscillations in the Solar Atmosphere (ROSA) imager at the Dunn Solar Telescope (DST).  The Hi-C mission \citep{kobayashi_2014} also reported coronal loop widths of $\sim200\;km$ (although not associated with a flare), using high resolution EUV observations (e.g. \citealt{peter2013,klimchuk2020}). These calculated coronal loop widths are all limited by the resolution of their respective telescopes.

As observations with the Daniel K. Inouye Solar Telescope (DKIST, \citealt{rimmele2020}) 4-meter-aperture telescope commence, we now have the capability to observe the Sun at unprecedented spatial scales with the Visible Broadband Imager (VBI, \citealt{woger2021}).  The pixel size of DKIST/VBI H$\alpha$ images is $0''.017$, half the diffraction limit of DKIST at this wavelength.  This provides a resolution at the solar surface of $\sim$$24\;km$, perhaps sufficient to resolve the elementary strands that constitute the arcade of PRFLs. 

Here we present the highest-resolution $H\alpha$ observations of a solar flare to date, observed with DKIST/VBI during the flare decay phase.  We specifically report on fine structure in on-disk, post-reconnection flare loops with structuring on scales roughly 3 to 4 times smaller than those previously reported with other observing facilities.  In Section \ref{sec:datasummary} we describe the DKIST, SDO, and GOES observations of the X1.3-class flare SOL2024-08-08T19:01:00.  In Section \ref{sec:methods} we describe our methods for analysis of this dataset.  In Section \ref{sec:finestructure} we present the results of a statistical study of the widths of $686$ PRFLs observed across the $20$ best-seeing images.  In Section \ref{sec:discussion} we discuss the implications of this study in the broader context of fine-structure observations and flare modeling.  In Section \ref{sec:conclusions} we summarize our main conclusions.

\section{Observations}\label{sec:datasummary}

\subsection{Summary of the X1.3-class flare on 8 August 2024: X-ray light curve and ribbon structure}

\begin{figure}
\includegraphics[width=1\textwidth]{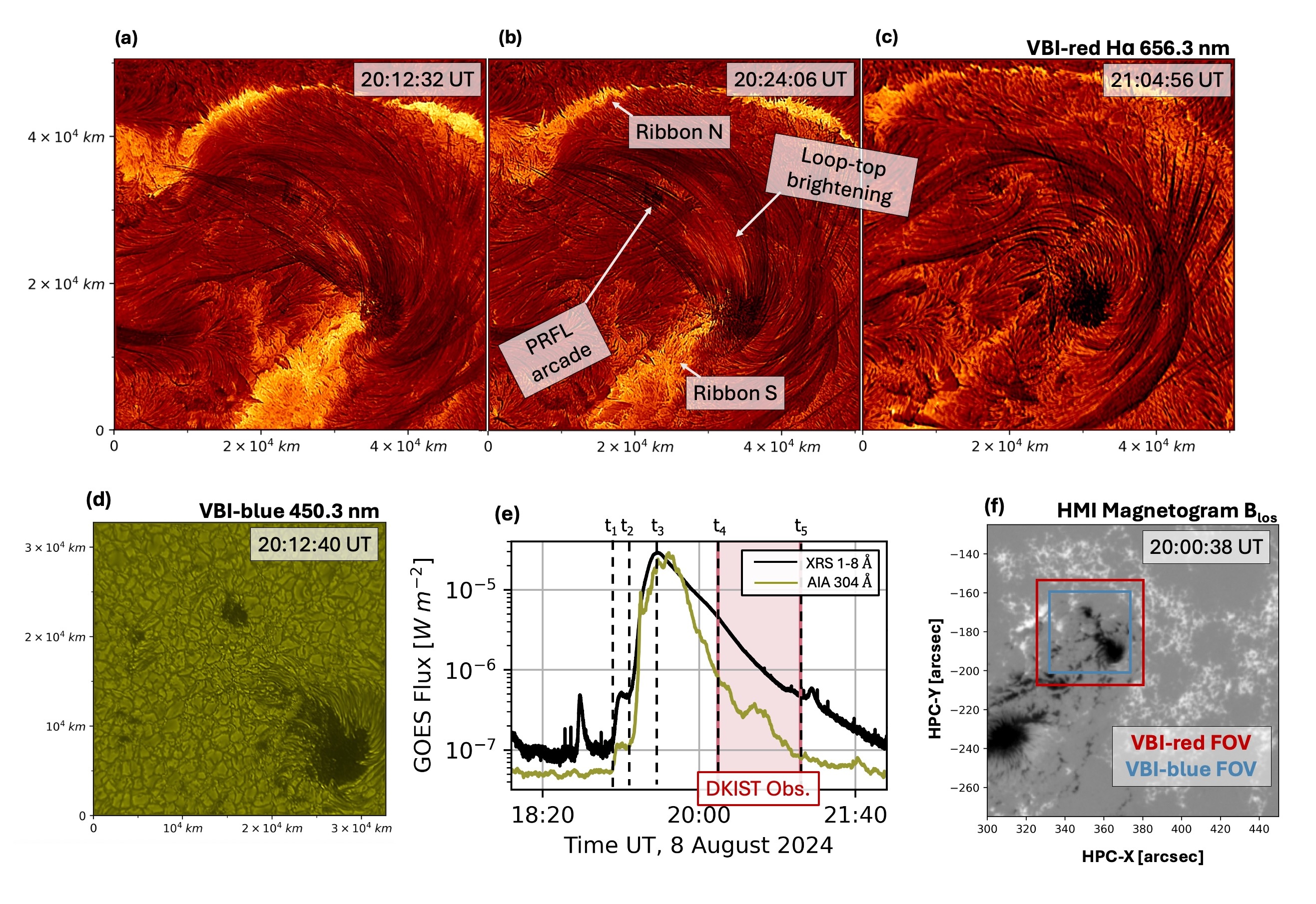}
\centering
\caption{Summary of the GOES/XRS X-ray class X1.3 flare on 8 August 2024.  (a)-(c) Images from the VBI-red H$\alpha$ 656.3 nm filter.  In (b) we indicate the locations of features discussed in this work, including ribbons N and S, the arcade composed mostly of dark coronal loops, and the loop-top brightening. In the supplementary material for this letter (Animation 1), we include a movie of the speckle-reconstructed, destretched VBI observations of this flare from 20:12 UT to $\sim$20:25 UT, prior to the loss of AO lock.  This animation includes the frames shown in panels (a) and (b) here. (d) Image from VBI-blue blue-continuum filter   centered at 450.3 nm at the start of DKIST observations. (e) GOES/XRS 1-8 \AA\ SXR and SDO/AIA 304 \AA\ light curves for the event.  The shaded region indicates the period of DKIST observations.  Dotted vertical lines indicate times of interest in the flare evolution, discussed in Section \ref{sec:datasummary}.  (f) SDO/HMI line-of-sight magnetogram immediately prior to DKIST observations for context, showing the VBI-red and VBI-blue fields-of-view in red and blue respectively.}
\label{fig:event_summary}
\end{figure}

A GOES X1.3 class flare occurred in active region (AR) NOAA 13777 on 8 August 2024 (start, peak, and end times 19:01UT, 19:35UT, and 19:57 UT respectively; identifier SOL2024-08-08T19:01:00). The flare occurred in two stages. In Figure \ref{fig:event_summary}(e) we indicate times of interest in the evolution of this flare, from $t_1$ to $t_5$.  Prior to $t_1$ there is a peak in the GOES/XRS light curve resulting from a brightening in a different active region. Within the field-of-view eventually covered by DKIST, there is a filament visible in the SDO/AIA 131 \AA\ and 304 \AA\ (not shown) at times $t_1$ and $t_2$. The first stage (smaller peak in the GOES XRS 1-8 \AA\ light curve between $t_1$ and $t_2$) coincides with a smaller reconnection event in loops connecting the region shown in the DKIST images to an eastward region. 

During the main reconnection event between $t_2$ and $t_3$, the filament mentioned above erupts, likely leading to the CME event listed in the LASCO catalog \citep{gopalswamy2009,gopalswamy2024}.  The peak of the GOES/XRS SXR light curve occurs at $\sim$19:35 UT ($t_3$).  The footpoints of the post-reconnection arcade after this eruption are visible as a southern ribbon (S) connecting two sunspots in AR13777 and an upper, arc-shaped ribbon (N) in the positive-polarity solar plage (see Figure \ref{fig:event_summary}(f) for the SDO/HMI magnetogram contextualizing these observations). In this letter we investigate detailed observations of flare loops that are present after the CME release, during the decay phase of the flare light curve between $t_4$ and $t_5$.

\subsection{High-resolution DKIST observations with the Visible Broadband Imager (VBI): Ribbons and PRFLs}\label{DKISTobs}

The DKIST observations of flare SOL2024-08-08T19:01:00 were collected in fulfillment of experiment ID 2.11.  The VBI was observing with the H$\alpha$ interference filter in the red channel (VBI-red, centered at 656.30 nm with an effective FWHM of $\sim$0.65\AA); DKIST data center product ID L1-ENORQ; $69''\times69''$ field-of-view, FOV; Figure \ref{fig:event_summary}(a)-(c)), and blue continuum in the blue channel (VBI-blue, centered at 450.3 nm with filter width 4.1\AA; DKIST data center product ID  L1-TQFBC; $45''\times45''$ FOV; Figure \ref{fig:event_summary}(d)).  Spatial samplings for VBI-red and VBI-blue images are $0''.017\;pix^{-1}$ and $0''.011\;pix^{-1}$, respectively.  The raw data have exposure times of 1 ms and 7 ms for the H$\alpha$ and blue continuum filters, respectively.  The VBI data were speckle-reconstructed \citep{woger2008}, with each reconstructed image (cadence of 2.66 s) produced from 80 raw images.  We destretched the image series to further account for atmospheric disturbances, but use the speckle-reconstructed, pre-destretched data during our analysis of PRFL fine structure below.  The mean Fried parameter for the dataset was $r_0=5.4\pm2.2 \; cm$, with a maximum value of $r_0 = 13.0 \; cm $ at 20:55 UT and several periods of successive image frames with $r_0>10\;cm$ (Table \ref{tab:framestats}).

Thin coronal loops connecting ribbons N and S are mostly in absorption in the VBI H$\alpha$ filter.  Each loop is nearly parallel to those in closest proximity, such that there is a clear arcade of coronal loops extending from ribbon N to ribbon S.  Around 20:20 UT, strands near the loop-top brighten individually as plasma cools through the temperatures to which H$\alpha$ is sensitive. In each loop, the region of locally brightened plasma begins near the loop-top and spreads outward in both directions along the loop, but the loop brightenings are not visible lower in the atmosphere (closer to the ribbons).  As the image sequence progresses, loops in the arcade are generally more perpendicular (less sheared) to the two ribbons.  Two small sunspots are visible beneath the arcade (clearest in the blue continuum image, Figure \ref{fig:event_summary}(d)). 

\section{Methods}\label{sec:methods}

To determine the cross-sectional widths of dark flare loops observed in the VBI H$\alpha$ images, we fit a negative-Gaussian empirical model to loops observed in images under the best seeing conditions.  A linear function is added to the Gaussian to account for gradients in the intensity of the bright chromospheric backdrop, against which the dark coronal loops are visible.  The model $L(x)$ is given by the following:

\begin{equation}
    L(x) = I_0e^{\frac{-(x-\mu_x)^2}{2\sigma^2}}+ax+b
\end{equation}

\noindent where $L(x)$ is the modeled intensity as a function of position $x$ along an empirically defined profile of the strand, $I_0$ is the amplitude (negative for the dark loops investigated here), $\mu_x$ is the central position of the model, $\sigma$ is the standard deviation of the model, and $a$, $b$ are parameters describing the linear fit to the bright background intensity.  We use the Python function \texttt{scipy.optimize.curve\_fit} to perform the fit.

Similarly to the determination of loop widths with the GST reported in \cite{jing2016}, we define the width $w$ of each loop to be the full width at half minimum (FWHM) of the fitted Gaussian profile, determined by:
\begin{equation}\label{eq:w}
    w=2\sigma\sqrt{2\ln2}
\end{equation}

We assume that the loops observed are locally cylindrical, such that projection effects from solar viewing angle do not affect our calculation.  As we discuss in Section \ref{sec:discussion}, coronal loops do not necessarily have a circular, nor uniform, cross-section.  We visually identify each loop and take particular care to select features that appear as single strands.  In a 3D scenario, many strands in the arcade may lie within the same line-of-sight but at different heights, below the angular separation necessary to resolve the features with VBI H$\alpha$.  We derive the error in cross-sectional width $\delta w$ by propagating the error in $\sigma$ ($\delta \sigma$, an output of the Python fitting function), from which $w$ was determined (Equation \ref{eq:w}).

\begin{equation}
    \delta w = |w\sqrt{(\delta \sigma/\sigma)^2}|
\end{equation}

\noindent The Fried parameters $r_0$ associated with the 20 best-seeing images are shown in Table \ref{tab:framestats}. We remove features with calculated error $\delta w > 15\;km$ in the cross-sectional width from our analysis.  This helps to reduce the effects of selection bias by (1) removing features that are likely composed of more than one strand; and (2) removing features that are too faint to reliably determine a width.  To verify our results, we perform our analysis several additional times on the five best-seeing image frames.  While there are small variations in the calculated widths from these runs, the major results remain the same.

In Figure \ref{fig:selfeats} we show the features selected using image Frame 3 at 20:12:40 UT.  We use the same three regions for analysis of each chosen frame.   We perform this analysis on the pre-destretched image sequence in order to avoid the influence of small-scale artifacts of the destretching process. 

In Section \ref{sec:symmetry} we describe the symmetry of the PRFL width distributions about their mean values. To test the symmetry of a distribution, we use (i) the Pearson median skewness, or second skewness coefficient, defined as $s=3*(mean-median)/std$, where values of $|s|<0.5$ indicate reasonable symmetry in the distribution and (ii) the Python function \texttt{scipy.stats.skewtest}, which tests the null hypothesis that the distribution from which a sample is drawn is the same as that of a corresponding normal distribution based on the Fisher-Pearson coefficient of skewness.  We note that failure to reject the null hypothesis is not definitive evidence of a symmetric distribution.

\section{Results}\label{sec:finestructure}
\subsection{Fine structure of coronal loops}\label{sec:loopwidths}

\begin{figure}
\includegraphics[width=.85\textwidth]{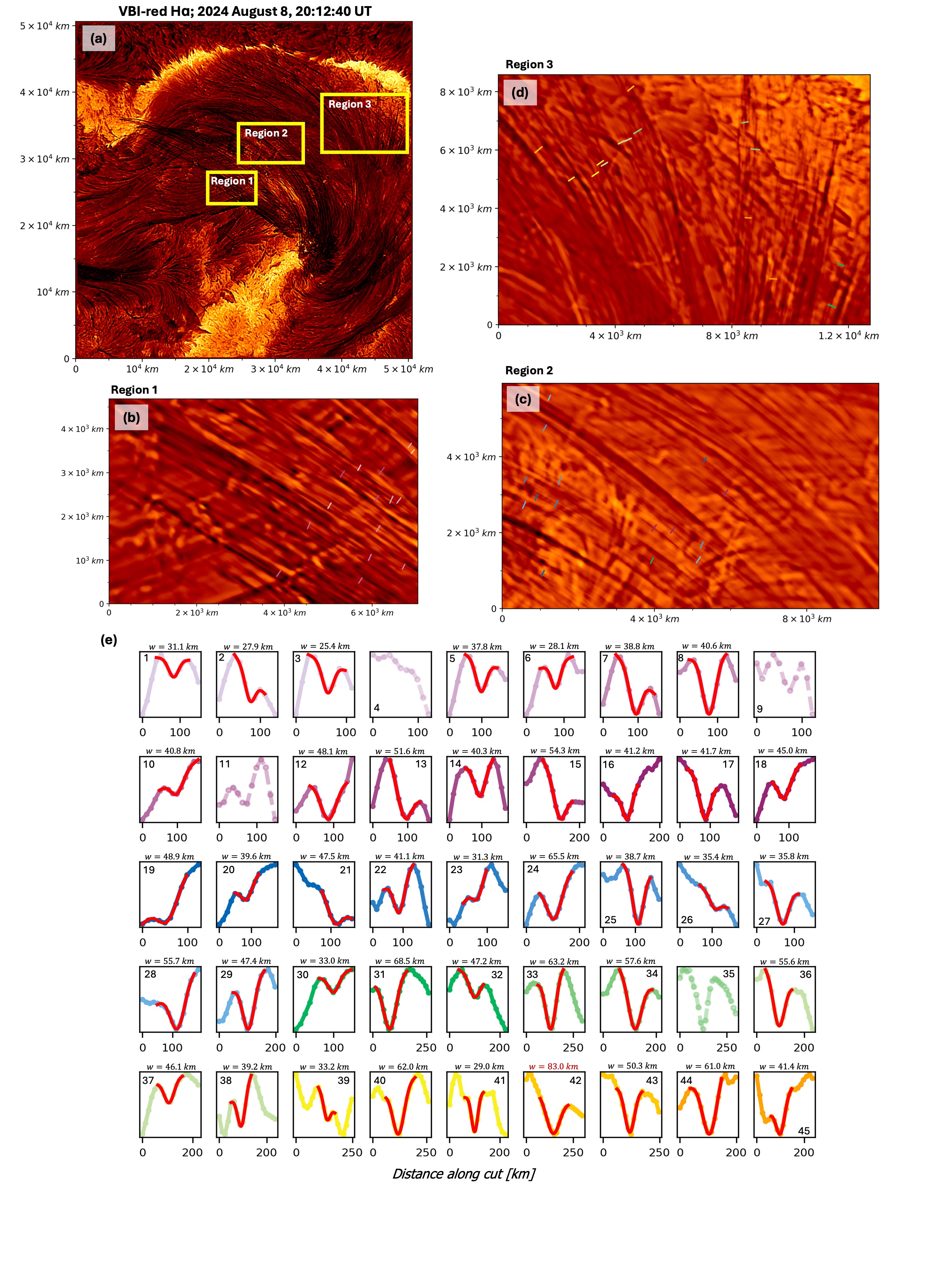}
\centering
\caption{Example of fine-scale coronal strands studied in this work.  (a) VBI-red $H\alpha$ image from 20:12:40 UT ($r_0 = 12.2\; cm$).  Yellow boxes indicate the three regions near the loop-top where we investigate the coronal strand width. (b)-(d) Regions of interest for loop width determination in Section \ref{sec:loopwidths}.  Colored line cuts are placed perpendicular to the thin strands investigated here. (e) Intensity profiles along the coronal strands indicated by the cuts in (b)-(d), and can vary between $\sim100\;km$ and $\sim250\;km$ in length. Dots on each profile are separated by $\sim12\;km$. Colors indicate the corresponding features in (b)-(d).  Red solid lines indicate a fit of the data to a function as in Equation (1), from which the FWHM loop width is determined.  Dashed intensity profiles without displayed fits in red indicate the features for which the algorithm could not find a successful fit. We note the derived widths of each feature.  Axis labels in (a)-(d) are in $km$ along the respective axes in the VBI images; in (e), axis labels are in $km$ along the perpendicular cuts.  }
\label{fig:selfeats}
\end{figure}

In Table \ref{tab:framestats} we show the results of the Gaussian-fitting width determination analysis for 45 PRFLs in each frame (3 selected windows with 15 features per window) in the 20 best-seeing frames (as determined by the Fried parameter) of the speckle-reconstructed VBI data between 20:12UT and 20:25UT, the first 13 minutes of observations during which the best seeing was achieved and AO lock was successful. The intensity profiles along the 45 chosen features in Frame 3 (20:12:40 UT), with model fits when appropriate, are shown in Figure \ref{fig:selfeats}(e).  The mean, minimum, and maximum loop widths are listed in Table \ref{tab:framestats}. 

\begin{figure}
\includegraphics[width=\textwidth]{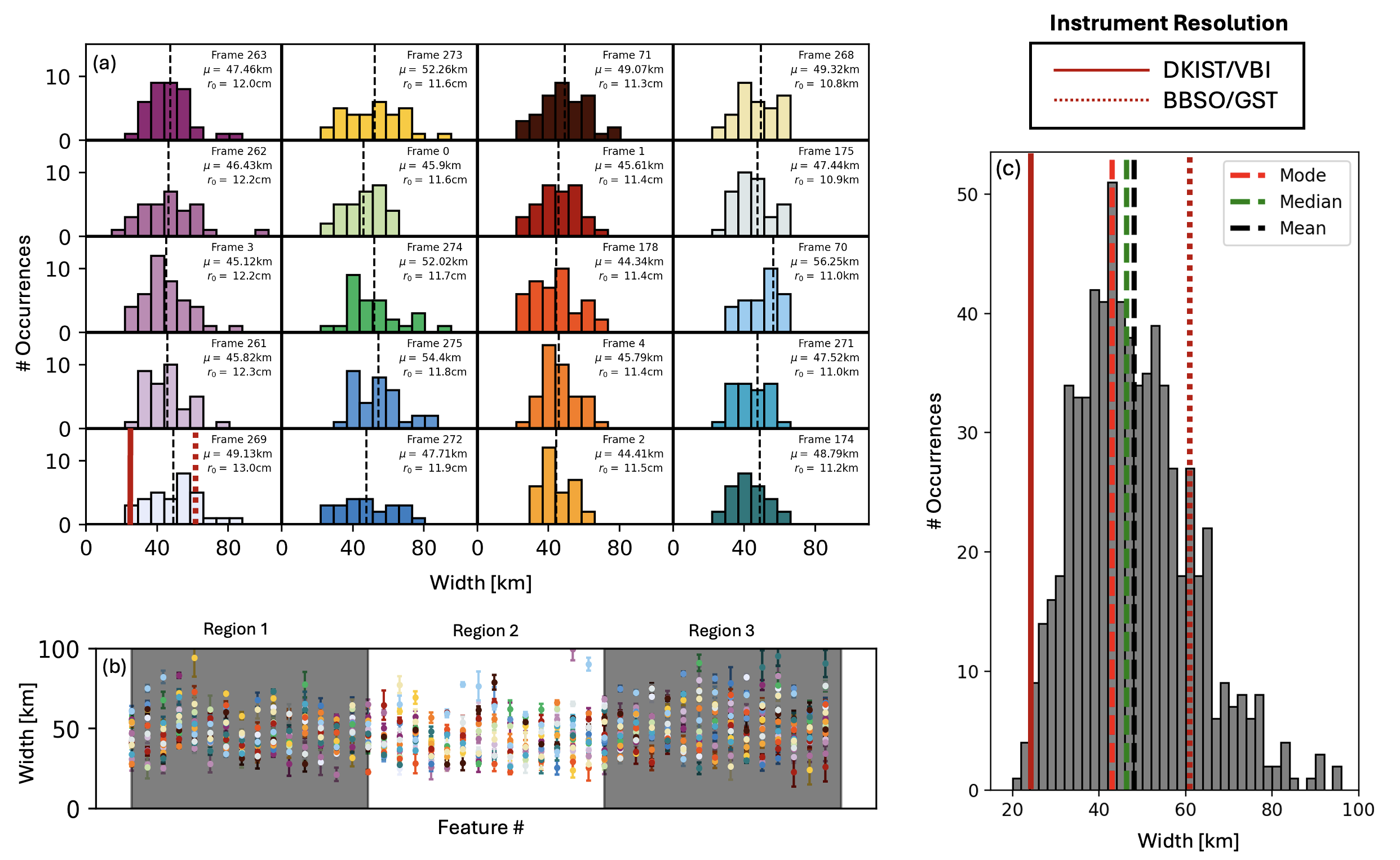}
\centering
\caption{Summary of loop widths in each of the 20 best-seeing image frames indicated in Table \ref{tab:framestats}.  (a) Histograms of the loop width distributions in each frame.  Colors do not correspond to those in Figure \ref{fig:selfeats}(e). Black dotted vertical lines indicate the distribution means. The third panel in the leftmost column is the distribution corresponding to the features shown in Figure \ref{fig:selfeats}. Vertical red lines in the bottom panel of the leftmost column indicate the spatial resolution of DKIST/VBI and BBSO/GST. (b) Loop widths with errors for all frames studied.  Alternating shaded regions indicate the features from each of the three regions in Figure \ref{fig:selfeats}, which are the same for all 20 frames. (c) Combined distribution of calculated widths from all 20 frames, with the mean and median of the distribution and the VBI and GST resolution limits indicated.}
\label{fig:features_widths}
\end{figure}

Of the 900 total features investigated across all 20 frames, 214 are poorly fit by the algorithm (indicating that the single-Gaussian fit is not appropriate for these features, most often because of overlapping strands within the line-of-sight) and are not included in the statistical analysis. Distributions of the cross-sectional widths of the loops observed in the 20 selected frames are shown in Figure \ref{fig:features_widths}(a). Figure \ref{fig:features_widths}(b) shows the individual calculated cross-sectional widths, with alternating shaded regions corresponding to the first, second, and third regions (from left to right) indicated by red boxes in Figure \ref{fig:selfeats}. The minimum calculated PRFL width is $21.4\;km$, essentially the diffraction limit of the telescope, and the mean widths of the distributions lie between $45$ and $55\;km$.  The minimum loop width lies below the $\sim$24 $km$ diffraction limit of DKIST in H$\alpha$; this results from our definition of PRFL width (Section \ref{sec:methods}) and, like in previous works (e.g. \citealt{schmidt2025}) is likely due to random errors in the Gaussian fitting process.

\begin{deluxetable*}{|c|c|c|c|c|c|c|c|c|}
\tabletypesize{\footnotesize}
\tablewidth{0pt}

 \tablecaption{Coronal loop width statistics \label{tab:framestats}}
 \tablehead{\colhead{\textbf{Frame}} & \colhead{\textbf{Obs. Time [UT]}} & \colhead{\textbf{$r_0$ [cm]}} & \colhead{\textbf{Min. Width [km]}} & 
 \colhead{\textbf{Max. Width [km]}} &
 \colhead{\textbf{Mean Width [km]}} &
 \colhead{\textbf{$\delta w$ [km]}} &
 \colhead{\textbf{Pearson's $s$}} &
 \colhead{\textbf{\texttt{skewtest}}}} 
\startdata
    0 & 20:12:32 & 11.6 & 25.9 & 62.5 & 45.9 & 3.3 & NS & S
    \\ \hline
    1 & 20:12:35 & 11.4 & 23.0 & 66.4 & 45.6 & 3.9 & S & S
    \\ \hline
    2 & 20:12:38 & 11.5 & 29.7 & 59.3 & 44.4 & 3.2 & NS & S
    \\ \hline
    3 & 20:12:40 & 12.2 & 25.4 & 83.0 & 45.1 & 2.8 & NS & NS
    \\ \hline
    4 & 20:12:43 & 11.4 & 28.1 & 66.4 & 45.8 & 3.1 & S & S
    \\ \hline
    70 & 20:15:39 & 11.0 & 30.9 & 90.2 & 56.3 & 3.3 & S & S
    \\ \hline
    71 & 20:15:42 & 11.3 & 27.8 & 79.0 & 49.1 & 3.4 & S & S
    \\ \hline
    174 & 20:20:16 & 11.2 & 25.2 & 95.3 & 48.8 & 3.8 & NS & NS
    \\ \hline
    175 & 20:20:19 & 10.9 & 27.0 & 76.8 & 47.4 & 3.6 & S & NS
    \\ \hline
    178 & 20:20:27 & 11.4 & 22.8 & 73.0 & 44.3 & 3.6 & S & S
    \\ \hline
    261 & 20:24:08 & 12.3 & 27.1 & 80.4 & 45.8 & 3.9 & S & NS
    \\ \hline
    262 & 20:24:11 & 12.2 & 21.4 & 99.7 & 46.4 & 4.3 & S & NS
    \\ \hline
    263 & 20:24:13 & 12.0 & 28.1 & 83.4 & 47.5 & 3.7 & NS & NS
    \\ \hline
    268 & 20:24:27 & 10.8 & 26.3 & 77.5 & 49.3 & 3.5 & S & S
    \\ \hline
    269 & 20:24:30 & 13.0 & 25.1 & 82.2 & 49.1 & 3.2 & S & S
    \\ \hline
    271 & 20:24:35  & 11.0 & 26.4 & 107.6 & 47.5 & 3.1 & NS & NS
    \\ \hline
    272 & 20:24:37 & 11.9 & 25.6 & 78.3 & 47.7 & 4.0 & S & S
    \\ \hline
    273 & 20:24:40 & 11.6 & 23.9 & 94.3 & 52.3 & 3.5 & S & S
    \\ \hline
    274 & 20:24:43 & 11.7 & 28.3 & 91.1 & 52.0 & 3.9 & NS & NS
    \\ \hline
    275 & 20:24:46 & 11.8 & 32.6 & 84.4 & 54.4 & 3.3 & S & S 
    \\ \hline   
\enddata
\tablecomments{Results from the 20 best-seeing image frames from the 8 August 2024 X1.2-class flare, in order of observation time.  In column 1 we indicate the frame number in the VBI data; in column 2 we list the observation time; in column 3 we list the Fried parameter; in columns 4 through 6 we list the minimum, maximum, and mean strand widths, respectively; in column 7 we list the average error in derived strand width; and in columns 8 and 9 we show skewness test results from tests (i) and (ii), where ``S" and ``NS" indicate that the condition for symmetry was met or was not met, respectively.}
\end{deluxetable*}

\subsection{Symmetry about the mean of PRFL width distributions}\label{sec:symmetry}

\cite{scullion2014} discuss that a representative sample of the true physical distribution of cross-sectional PRFL widths should be Gaussian or otherwise symmetric around the peak ``typical" strand width. However, distributions of these widths from other observatories have often shown an exponential increase in the number of features at the finest scales resolvable by those instruments (e.g. \citealt{scullion2014}).  For higher-resolution observations, for example with the GST, distributions are more peaked with tails at larger and smaller scales, although still with a longer tail at larger loop widths \citep{jing2016,schmidt2025}. The asymmetry indicates that we have until now been unable to produce a distribution of strand cross-sectional widths reflecting the fundamental loop widths.  
In this dataset, the distributions of loop widths from individual frames (Figure \ref{fig:features_widths}(a)) are often symmetric.   Figure \ref{fig:features_widths}(c) shows the distribution of loop widths across all 20 frames, with median $46\;km$ and mean $48\;km$.  The results of tests of symmetry (i) and (ii) for each distribution are listed in the last two columns of Table \ref{tab:framestats}.  13 of 20 and 12 of 20 distributions pass the first and second tests of symmetry, respectively.  The distribution of all widths in Figure \ref{fig:features_widths}(c) passes the first test of symmetry ($s\approx0.4$) but not the second. As a caveat, some distributions that do not pass one or both tests (e.g. corresponding to Frame 3 (Figure \ref{fig:selfeats}) and the combined distribution in Figure \ref{fig:features_widths}(c)) are penalized significantly for a few strands with widths $\sim$75\% greater than the mean width of the distribution.  The strands with large calculated widths may in reality be structures composed of many strands within the same line-of-sight (e.g. strand number 42 in the case of Frame 3, with width highlighted in red in Figure \ref{fig:selfeats}(e)).  Removal of these specific structures from the samples changes the result of the tests to favor symmetry, possibly including the combined distribution (i.e. our determination of whether a distribution can be considered symmetric is conservative).

\section{Discussion}\label{sec:discussion}

The $\sim20-80\;km$ widths of resolved coronal magnetic structures reported in this study have major implications in our understanding of coronal evolution during a flare.  For example, the mechanism by which the cross-section of a loop expands with height into the corona (due to the decrease of magnetic pressure with height and the requisite conservation of magnetic flux) is physically justified, but the expansion of coronal loops has never been clearly observed (e.g. \citealt{klimchuk2001,klimchuk2020,reep2022}). In the absence of decisive observations, this fundamental assumption has been challenged. In an observational study, \cite{klimchuk2001} suggested that the cross-sectional widths of coronal loops do not vary along their length, while  \cite{peter2012} argue that the interpretation of this analysis may be more complicated.  Due to the response functions of observed lines, only a narrow portion of the imaged coronal loop might be within the range of temperatures and densities to which any specific line is sensitive, giving the impression of a loop of constant width from photosphere to corona.  \cite{mikic_2013} note that cross-sectional expansion should be an assumed property of loop models, rather than an optional parameter, and demonstrate that processes such as coronal rain require geometrical non-uniformity.  

Besides the loop expansion, the shape and eccentricity of the loop cross-section introduce another set of important parameters for both PRFLs and loops not associated with a flare (e.g. \citealt{antiochos_1976,schrijver1989,klimchuk2001,vanBallegooijen2011,klimchuk2020} and references therein), although there may be relevant physical differences between flare-related and non-flare loop geometry.
\cite{klimchuk2020} use Hi-C observations of 20 coronal loops to argue that observed flux tubes are approximately circular, but that the unresolved loops may be composed of many strands of nonuniform cross-section.  Our observations align with this concept: the flare loops measured in our work are more than an order of magnitude smaller than the loops resolvable in Hi-C observations, suggesting that large, under-resolved, measurably-circular structures in past works could potentially be composed of many smaller $\sim20-80\;km$ wide loops.  Even the largest loop observed by the DKIST in this study is smaller than the vast majority of loops recorded by BBSO/GST, the telescope with the next-best resolving power (e.g. \citealt{jing2016,schmidt2025}).  A future study could investigate flare loops observed by both instruments simultaneously, and test whether the larger structures measured by previous works are composed of smaller loops resolvable by the DKIST.  

Properly constrained loop width is also relevant when considering the processes associated with flare heating, both in the corona and lower atmosphere.  For example, \cite{qiu2016} explain the long duration of gradual flare heating in SDO/AIA observations channels by assuming that each loop observed by SDO is composed of many individually heated strands, although the number of strands is undetermined. \cite{litwicka_2025} use the FLARIX code to address the remaining discrepancy between the modeled and observed intensities of chromospheric lines by assuming flare heating is provided by short pulses in unresolved filament structures. \cite{reep2018} use HYDRAD to investigate the impact of multi-threaded heating on synthetic \ion{Si}{IV} and \ion{Fe}{XXI} lines, assuming a uniform loop cross-section.  However, these works necessarily assume or infer coronal filling factors and relative loop widths.  The resolving power of the DKIST provides a valuable constraint on the loop width parameter, which may significantly improve such models.  Additionally, the relevance of spatially resolved flare loops and implications in flare heating go beyond the Sun; even in the case of stellar flares, heating of coronal loops associated with a flare has recently garnered attention as a potential mechanism for observed late-phase emission (e.g. \citealt{heinzel2018,yang2023,otsu2024})

Beyond the typical strand size, accurately constrained loop geometry is relevant to modeling the hydrodynamics (and even radiative-hydrodynamics) of a flare (e.g. \citealt{mikic_2013,reep2022,Reep_Airapetian_2023,reep2024,kowalski2024}), as discussed in Section \ref{sec:intro}. Even without the aid of these loop models, \cite{emslie_1994} performed an early study investigating the asymmetry of flux tubes.  Without sufficient resolving power, it is impossible to observationally determine the relevant parameters (such as cross-sectional area and eccentricity) to validate these experiments.  DKIST also provides the opportunity for a new multi-filter investigation of observed loops beyond just H$\alpha$ (e.g. in \ion{Ca}{II}~K) in order to constrain PRFL geometry in comparison to modern loop models. Even if the temperature response function of an individual filter prohibits single-line determination of the relevant geometrical properties (as suggested by \citealt{peter2012}), simultaneous observation of elementary structures with many filters could place additional constraints on hydrodynamical models, potentially presenting a solution to the problem at unprecedented scales.

Even with the high resolution of DKIST/VBI, it is unclear whether the actual characteristic loop widths might be even smaller than the diffraction limit of the DKIST. The symmetry argument in Section \ref{sec:symmetry}, which demonstrates that a distribution of loop widths measured with the VBI under ideal seeing conditions is often centered around the typical loop width, presents evidence that the DKIST is capable of resolving most elementary H$\alpha$ loops. We note that the distribution of loop widths produced by \cite{schmidt2025} using higher-resolution GST observations is right skewed, but is still more symmetric than distributions produced using lower-resolution observations (e.g. by the Swedish Solar Telescope as in \citealt{scullion2014}).  We note that the GST observations were taken of off-limb loops rather than the dark on-disk structures we study, and so the two may not be directly analogous.  However, the similarity between the two produced distributions presents the possibility that the typical loop widths observed with VBI may still lie above a characteristic width, albeit at higher resolution than GST.  To test the possibility that DKIST can observe most fundamental PRFLs more rigorously, the spectral and thermodynamic properties of the observed strands could be compared to realistic hydrodynamic simulations like HYDRAD  (e.g. spectral line intensity and loop cooling time, as in \citealt{reep2022}). Although these models are 1D, the magnetic field strength and other properties can be varied in these models to reflect the practical impacts of geometrical variation. If the evolution of the modeled and observed loops are relatively in agreement, this would provide further evidence to support our result.  Comparison to future 3D models or mathematical predictions of expected loop width (e.g. \citealt{peter2013}) could also be employed.

We note also that smaller coronal loop widths than we observe here have been estimated from mathematical arguments in the past.  Studying the same flare as \cite{klimchuk2020}, \cite{peter2013} estimate that coronal loops observed near the chromosphere may be composed of strands with widths far smaller ($\sim 15\;km$) than the $\sim45 \;km$ cross-sectional loop width that we claim using the VBI data.  They use 193\,\AA\ images with the Hi-C rocket for their analysis and note that loop cross-sections will vary based on the wavelength of observations and the filling factor of bright coronal plasma.  The maximum estimated width in H$\alpha$ may therefore differ.  We also note that the loops studied by \cite{peter2013}, while observed in an active region, were not associated with a flare. The high temperatures associated with coronal flare loops could also potentially impact the predicted loop widths.  Additionally, their analysis does not address the important effects of loop expansion or non-circular cross-section.  Other studies (e.g. \citealt{brooks2012}) estimate larger strand widths, indicating that the question of fundamental strand width can only be directly resolved through observations.

The data studied by \cite{jing2016} were taken using a Fabry-Perot-based, 0.07\AA\ FWHM filter centered at H$\alpha$$\;+\;1.0$\AA, while the DKIST observations discussed here are taken at line center with a filter width of $\sim$0.65\AA.  This detail may account for some difference in the measured loop widths with these respective telescopes, although a precise investigation of this impact will require an analysis of full spectral profiles, or the derivation of temperatures and densities from loop modeling, coupled with a full 3D radiative transfer treatment of H$\alpha$ (e.g. \citealt{leenaarts_2012,bjorgen2019}).  We note that the loops are dark in the H$\alpha$ images from both telescopes, indicating there is a net absorption in both passbands, which would be due to a combination of a changed temperature, density, populations, and velocity of the plasma in these loops. The broader VBI filter shows more contributions from the photospheric layers in the wings of the profile, resulting in some faint granulation visible in Figure \ref{fig:event_summary}(a)-(c)). Since the identification of coronal loops relies upon the contrast with respect to other nearby emission, the enhanced photospheric background with an interference filter may make it more difficult to identify PRFLs with DKIST when compared to GST. However, since the loops observed with the two filters are at least morphologically similar, and both resolve down to the diffraction limit of their respective telescopes, the most likely explanation for the differences in measured loop width is that the PRFLs observed by DKIST are simply unresolved by GST.

Whether or not the DKIST is capable of observing a large number of elementary coronal loops, the flare observations here have highlighted complications in the pursuit of observationally characterizing loop shape and evolution, including seeing, temperature sensitivity of the observed line, and projection effects.  Many of the coronal loops observed in our data are only visible in H$\alpha$ for a portion of their extent, possibly supporting the notion that temperature variations along a loop have a significant impact on our ability to investigate coronal loop geometry - but we note that the H$\alpha$ strength in these loops may be sensitive to many other physical parameters.  It is also possible that the distributions of loop widths shown in Figure \ref{fig:features_widths}, while more often symmetric than those presented in previous works, may still be somewhat limited by the resolution of the DKIST at the finest scales.  To validate and expand these results, the analysis should be performed for more events and for other spectral passbands. 

\section{Conclusions}\label{sec:conclusions}

The geometry of coronal loops has significant implications in many aspects of solar physics, including detailed modeling of post-reconnection flare loop thermal evolution and coronal structure in the absence of a flare.  However, the widths of elementary loops, theorized at $\sim10-200$~km, have historically been below the resolution limits of modern telescopes.  The 4-meter Daniel K. Inouye Solar Telescope has the capability to observe structures less than half the size of the smallest-observable features in the past, and could possibly resolve a significant portion of elementary coronal loops.

We use DKIST/VBI H$\alpha$ observations of the decay phase of the X1.3-class flare SOL2024-08-08T19:01:00 to analyze fine structure of post-reconnection flare loop strands and compare these with previous observations.  Our major findings are as follows:

\begin{enumerate}
    \item We find that DKIST observations reveal fine structure at or near the diffraction limit of the DKIST in the flare loops and flare ribbons. Observations of this flare at such high resolution and cadence are the first of their kind and provide fertile ground for future investigations of the solar flare dynamics.
    \item We measure the widths of 686 post-reconnection flare loops across the 20 best-seeing frames and find average loop widths of $44\; km < w < 56\; km$, with minimum loop widths at the diffraction limit ($\sim$$24\;km$ in H$\alpha$).  On average, these strands are $\sim$45\% smaller than the next-smallest observed H$\alpha$ PRFLs captured by the Goode Solar Telescope at BBSO \citep{jing2016}.
    \item According to two tests of the symmetry of loop width distributions, we find that 13/20 and 12/20 of the distributions pass tests of symmetry with statistical significance.  The symmetry of loop distributions provides preliminary evidence that the DKIST/VBI can resolve most elementary flare loops.  
    \item From the discovered symmetrical loop width distributions, we conclude that previous studies of coronal loop geometry have been limited by insufficient telescope resolution.  This result suggests that the DKIST data could provide valuable constraints on modern hydrodynamic and radiative-hydrodynamic flare models.
\end{enumerate}

This is the first high-resolution solar flare study with DKIST/VBI in H$\alpha$.  The resolution achieved by the DKIST demonstrates significant substructure well below the scales previously observable, indicating that it is now both possible and necessary to revisit past models of arcade structure in the context of high-resolution DKIST observation.  Since both the dark loops in the post-eruption arcade and bright flare ribbons are visible in H$\alpha$, a future study may investigate the common spatial scales between these two regions, testing the connectivity between sites of chromospheric energy deposition and the coronal loop structures resulting from magnetic reconnection.  We anticipate that these observations will represent the first of many from the DKIST to be used to investigate the role of fine structure in flare heating, placing additional constraints on modern simulations and contributing to a detailed picture of flare evolution in the chromosphere and corona.
     
\section{Acknowledgements}

The research reported herein is based in part on data collected with the Daniel K. Inouye Solar Telescope (DKIST) a facility of the National Science Foundation (NSF).  DKIST is operated by the National Solar Observatory (NSO) under a cooperative agreement with the Association of Universities for Research in Astronomy (AURA), Inc.  DKIST is located on land of spiritual and cultural significance to Native Hawaiian people. The use of this important site to further scientific knowledge is done with appreciation and respect.  Support for this work is provided by the NSF through the DKIST Ambassadors program, administered by NSO and AURA, Inc.  M.D.K. acknowledges support by NASA ECIP award 80NSSC19K0910 and NSF CAREER award SPVKK1R
C2MZ3. R.J.F. thanks support from NASA HGI award 80NSSC25K7927. K.P.R. acknowledges support from NASA award 80NSSC20K1282.

\bibliography{bib}
\bibliographystyle{aasjournal}

\end{document}